# An electrochemical thermal transistor

Aditya Sood [1,2], Feng Xiong [1,3,9], Shunda Chen[4], Haotian Wang [1,10], Daniele Selli[5,11], Jinsong Zhang[1], Connor J. McClellan[3], Jie Sun[1,12], Davide Donadio[4,6], Yi Cui[1,7], Eric Pop [1,3,8] & Kenneth E. Goodson[1,2]

The ability to actively regulate heat flow at the nanoscale could be a game changer for applications in thermal management and energy harvesting. Such a breakthrough could also enable the control of heat flow using thermal circuits, in a manner analogous to electronic circuits. Here we demonstrate switchable thermal transistors with an order of magnitude thermal on/off ratio, based on reversible electrochemical lithium intercalation in $MoS_2$ thin films. We use spatially-resolved time-domain thermoreflectance to map the lithium ion distribution during device operation, and atomic force microscopy to show that the lithiated state correlates with increased thickness and surface roughness. First principles calculations reveal that the thermal conductance modulation is due to phonon scattering by lithium rattler modes, *c*-axis strain, and stacking disorder. This study lays the foundation for electrochemically-driven nanoscale thermal regulators, and establishes thermal metrology as a useful probe of spatio-temporal intercalant dynamics in nanomaterials.

[1] Department of Materials Science and Engineering, Stanford University, Stanford, CA 94305, USA. [2] Department of Mechanical Engineering, Stanford University, Stanford, CA 94305, USA. [3] Department of Electrical Engineering, Stanford University, Stanford, CA 94305, USA. [4] Department of Chemistry, University of California, Davis, CA 95616, USA. [5] Max Planck Institute for Polymer Research, Ackermannweg 10, D-55128 Mainz, Germany. [6] Ikerbasque, Basque Foundation for Science, E-48011 Bilbao, Spain. [7] Stanford Institute for Materials and Energy Science, SLAC National Accelerator Laboratory, Menlo Park, CA 94025, USA. [8] Precourt Institute for Energy, Stanford University, Stanford, CA 94305, USA. [9] Present address: Department of Electrical and Computer Engineering, University of Pittsburgh, Pittsburgh, PA 15261, USA. [10] Present address: Department of Chemical and Biomolecular Engineering, Rice University, Houston, TX 77005, USA. [11] Present address: Dipartimento di Scienza dei Materiali, Universita di Milano-Bicocca, 20125 Milano, Italy. [12] Present address: School of Chemical Engineering and Technology, Tianjin University, 300350 Tianjin, China. These authors contributed equally: Aditya Sood, Feng Xiong. Correspondence and requests for materials should be addressed to Y.C. (email: yicui@stanford.edu) or to E.P. (email: epop@stanford.edu) or to K.E.G. (email: goodson@stanford.edu)

Achieving dynamic control of heat flow at the nanoscale represents an outstanding challenge in engineering energy transport. A thermal transistor, a device whose thermal conductance can be modulated in real-time using an external stimulus, has the potential for transformative applications in dynamic thermal management[1], energy harvesting[2], and phonon logic[3,4]. For decades, there have been numerous studies of fundamental relationships between material microstructure and thermal conductivity. For example impurities[5–7], vacancies[8], dislocations[9], grain boundaries[10–12], and interfaces[13–15] can all have a significant impact on the thermal conductivity. Despite progress on the fundamental understanding of these relationships, there are very few demonstrations of tuning nanoscale thermal transport in real-time. Dynamic changes in microstructure have been leveraged to actively modulate the thermal conductivity of a material[16], e.g., by temperature-induced phase change[17,18], electric field[19], and electrochemical intercalation[20,21]. However, most prior demonstrations of reversible tuning at room temperature have shown relatively small thermal on/off ratios, below approximately 1.6×, with switching time scales on the order of hours[20,21]. Practical applications of thermal switches would require devices with significantly larger on/off ratios, and nanoscale dimensions in order to achieve faster operation.

Here we demonstrate reversible thermal conductance modulation in a fully dense medium by a factor of nearly 10×, on time scales of minutes, across a material that is only 10 nm thick. The thermal transistors are based on nanoscale $MoS_2$ films actuated by reversible electrochemical intercalation of Li ions, with the on-state corresponding to the pristine $MoS_2$ and the off-state corresponding to the Li-intercalated $MoS_2$. Thermal conductance measurements are performed in situ using time-domain thermoreflectance (TDTR) while the device (which operates like a $MoS_2$ nanobattery) undergoes reversible electrochemical cycling. Using operando TDTR microscopy, we also probe the spatial distribution of Li ions inside $MoS_2$ at different stages of intercalation within an electrochemical cycle. Correlating atomic force microscopy (AFM) with thermal measurements reveals a strong impact of Li on microstructure in terms of c-axis lattice expansion, and significant mesoscopic disorder. Ab initio density functional theory (DFT) and non-equilibrium molecular dynamics (NEMD) calculations sort out the relative contributions of the different factors leading to the thermal conductance modulation.

## Results
**Device preparation.** Figure 1 shows a schematic of the transparent electrochemical cell[22] used for operando thermal conductance measurements (see Methods for further details of device fabrication and packaging). The working device is a 10 nm thick layered $MoS_2$ crystal, prepared on $SiO_2$ (90 nm) on p-type Si. An 80 nm Al layer patterned on top of the $MoS_2$ serves as an electrical contact and as opto-thermal transducer for TDTR measurements. A solid Li pellet acts as the reference and counter electrodes, and 1.0 M $LiPF_6$ in ethylene carbonate/diethyl carbonate (EC/DEC, 1:1 w/w) is the liquid electrolyte. TDTR is an optical pump-probe technique that is used to measure cross-plane thermal transport in thin-films (see Methods, Supplementary Figs. 1–4, Supplementary Table 1). Because it is an optical method, TDTR is well-suited for non-invasive thermal measurements of such devices during electrochemical operation. We perform real-time measurements by fixing the pump-probe delay time to +100 ps and continuously collecting the in-phase and out-of-phase voltages ($V_{in}$ and $V_{out}$). This enables measurement of thermal conductance as a function of both time and spatial coordinate. To interpret the measurement, we fit the effective cross-plane thermal conductance of $MoS_2$ ($G$) such that $1/G$ represents the combined intrinsic resistance of the film plus the Al/$MoS_2$ and $MoS_2$/$SiO_2$ interfaces acting in series.

**Operando thermal conductance microscopy.** It is generally known that electrochemical intercalation in single particles can result in an inhomogeneous Li ion distribution. A few recent studies have used in situ optical and spectroscopic techniques to visualize this inhomogeneity at the nanoscale[22,23]. Here we effectively probe Li ion segregation in single crystal $MoS_2$ films through its impact on the local cross-plane thermal transport. This represents a demonstration of operando thermal conductance microscopy performed during intercalation of a single particle battery electrode.

To probe this spatial distribution, we use a modification of the traditional TDTR technique[12]. The sample (see optical image in Fig. 2a) is placed on a two-axis translation stage, allowing it to be raster scanned in the plane perpendicular to the laser's optical axis. Measurements are performed under equilibrium electrochemical conditions in potentiostatic mode, with the potential of the $MoS_2$ working electrode $V_{WE}$ (relative to Li$^+$/Li) fixed at values ranging between the strongly lithiated and delithiated states at 1.0 V and 3.0 V, respectively. We do not use voltages much lower than 1.0 V to avoid the irreversible formation of products such as $Li_2S$ and Mo[24]. Figure 2b shows thermal conductance maps as the device is sequentially subjected to voltages from 1.8 V to 1.0 V (lithiation), and 1.2 V to 3.0 V (delithiation). A video montage of these images is provided in Supplementary Movie 1.

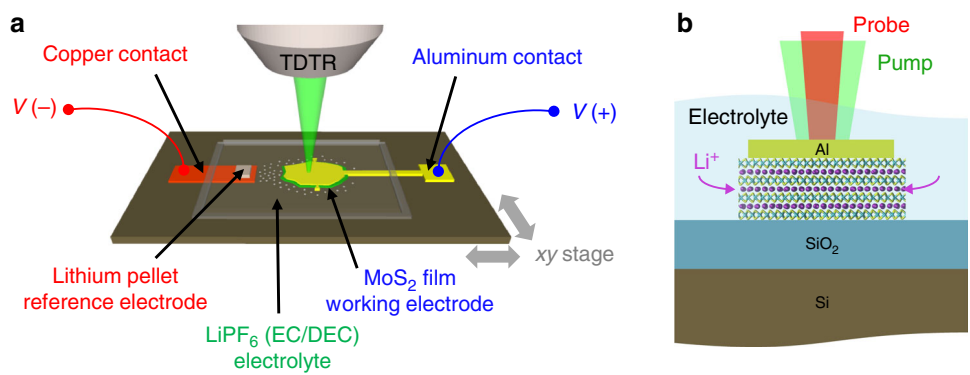

**Fig. 1** Experimental measurement of a thermal transistor. **a** Schematic of the electrochemical cell used for operando TDTR experiments. **b** Cross-sectional view of a device under operation. Li ions enter and leave the $MoS_2$ film through the exposed edges

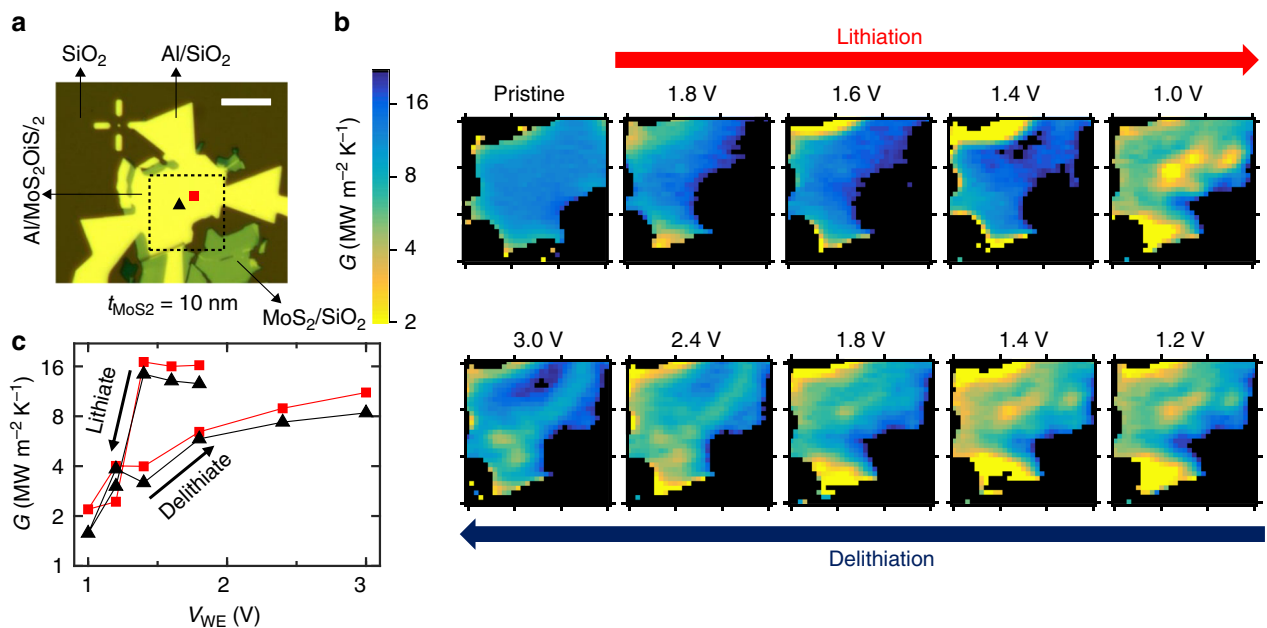

**Fig. 2** Operando scanning thermal conductance imaging. **a** Optical micrograph of the 10 nm thick MoS$_2$ device. Thermal conductance images are measured within a 15 × 15 μm square region marked by the dotted lines. The scale bar is 10 μm. **b** Maps of the inhomogeneous thermal conductance within the device taken at different stages of lithiation and delithiation over one electrochemical cycle. These are measured after holding the MoS$_2$ device at a constant potential $V_{WE}$ (relative to Li$^+$/Li) ranging from 1.8 V to 1.0 V for discharging (lithiation) and 1.2 V to 3.0 V for charging (delithiation). **c** Single-point thermal conductance vs. voltage, tracked over two spots that are indicated in **a** by the black triangle and red square

As the device is gradually lithiated from 1.8 V to 1.4 V, the images reveal regions of low thermal conductance $G$ forming at the top- and bottom-left edges of the device. We suggest that this is due to an increased concentration of Li near the exposed edges of the MoS$_2$. At $V_{WE}$ = 1.0 V, we observe the formation of a region of low $G$ at the center of the device. As $V_{WE}$ is raised from 1.2 V to 3.0 V, Li ions leave the film, resulting in a gradual dissolution of the low conductance region at the device center. By the end of the delithiation step at 2.4 V and 3.0 V, we see some spots of low $G$ forming in the lower half of the device. These results reveal that Li intercalation in thin-film MoS$_2$ likely occurs through the formation of Li-rich domains, qualitatively similar to the observations by Lim et al.[23] in single micro-platelets of LiFePO$_4$. In Fig. 2c, we track the evolution of $G$ at two locations (indicated in Fig. 2a) over the full lithiation-delithiation cycle. We find that $G$ decreases with decreasing $V_{WE}$ during the lithiation phase, and increases with increasing $V_{WE}$ during the delithiation phase.

**Dynamic thermal conductance modulation.** Real-time modulation of thermal conductance is achieved by charging and discharging the cell at constant current (i.e., in the galvanostatic mode), while simultaneously measuring $G$ at a single point on the device, shown in Fig. 3a. We apply a current of −1.2 nA and +1.2 nA to lithiate and delithiate the thermal transistor, respectively, while fixing the voltage limits for $V_{WE}$ at 1.0 V and 2.9 V. Note that since the Al metal electrode contacts multiple MoS$_2$ flakes, only a portion of the supplied current flows through the device under study. The full cycle time is 14 min, corresponding to a charge and discharge rate of about 8.5 C. Figure 3b shows a plot of $V_{WE}$ vs. time, and Fig. 3c shows the corresponding $G$ measured at the location indicated by the blue circle in Fig. 3a. During the discharge step, a negative current lithiates MoS$_2$ and decreases its thermal conductance from the pre-lithiation value of 15 ± 2 MWm$^{-2}$K$^{-1}$ to 1.6 ± 0.3 MWm$^{-2}$K$^{-1}$ (see Supplementary Fig. 3 for error analysis). When the current is reversed during the charge step, Li ions are removed from the MoS$_2$ and its conductance recovers to its pre-lithiation value. We measure a thermal conductance on/off ratio of 8–10× between the delithiated and lithiated states, which to the best of our knowledge is higher than that observed in previous in situ experiments on nanoscale thermal devices. A schematic of our electrochemically-gated thermal transistor is shown in Fig. 3d.

Figure 3e displays a plot of measured $G$ vs. $V_{WE}$, showing significant hysteresis between the charge and discharge curves. This is due to a hysteresis in the voltage curves themselves, and can be seen when we plot the average lithium composition $\chi$ (in Li$_\chi$MoS$_2$) vs. $V_{WE}$. Voltage curve hysteresis is a common feature in several battery electrode systems, and it is especially prominent in this case due to the large electrode size and relatively high charge and discharge rates[25]. To estimate $\chi$, we assume that the sample is fully delithiated at $V_{WE}$ = 2.9 V ($\chi$ = 0) and fully lithiated at $V_{WE}$ = 1.0 V ($\chi$ = 1)[26,27] (see Supplementary Fig. 5), varying linearly with intercalation time for intermediate voltages. When we plot thermal conductance vs. the average Li composition determined in this approximate manner (Fig. 3f), we find reduced hysteresis, suggesting a direct physical link between the two quantities.

**Ex situ measurements on chemically intercalated MoS$_2$.** In addition to operando studies, we also study the impact of chemical lithiation on thermal conductance of MoS$_2$ through ex situ measurements. Layered, crystalline MoS$_2$ films were exfoliated onto SiO$_2$ (90 nm) on p-type Si substrates. The samples were transferred into an Ar-filled glove box, and immersed in a 1.6 M n-Butyllithium in hexane solution (Fisher Scientific) for 2 h at 295 K. They were gently washed with anhydrous hexane to remove organic residues, and allowed to dry. The samples were sealed inside an airtight pouch and transferred to an e-beam evaporator for blanket deposition of the 81 nm thick Al transducer for TDTR measurements. These samples are thicker than

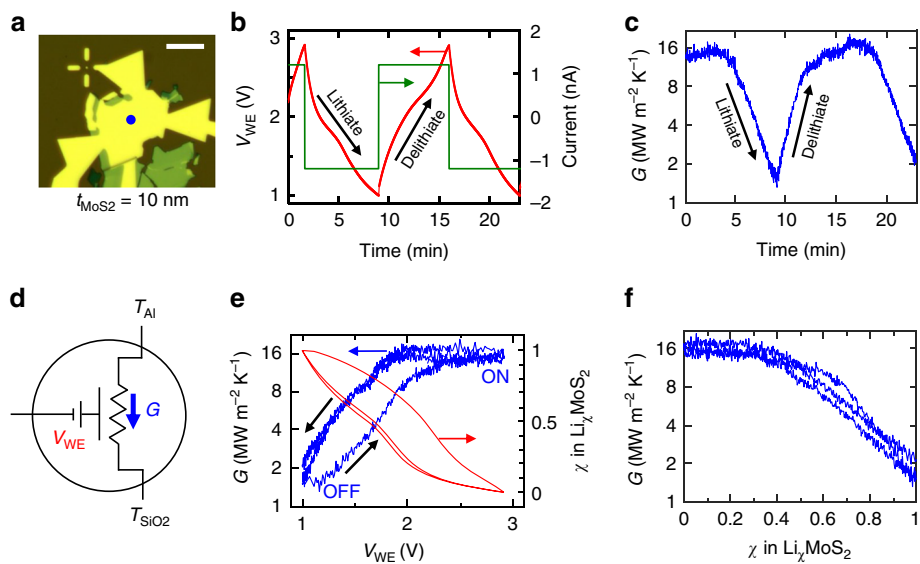

**Fig. 3** Thermal transistor characteristics. **a** Optical micrograph of the device showing the location of real-time TDTR measurements (blue circle). The scale bar is 10 μm. **b** Galvanostatic characteristics, obtained using an applied (dis)charge current of (−)+1.2 nA (shown in green). The resulting voltage curves are shown in red, taken within fixed limits of 1.0 V and 2.9 V. **c** Cross-plane thermal conductance measured during the electrochemical cycle shown in **b**. **d** Circuit diagram of the thermal transistor device: the gating voltage $V_{WE}$) that is applied between Li and Al/MoS$_2$ electrodes changes the thermal conductance G) to heat flow between the Al transducer and SiO$_2$ substrate, which are at temperatures $T_{Al}$ and $T_{SiO2}$ respectively. **e** Thermal conductance and average lithium composition $\chi$ plotted vs. voltage, showing significant hysteresis between charge and discharge curves. **f** G plotted against $\chi$

the device used for operando measurements, in the 60–100 nm range.

Figure 4a shows a thermal conductance map of a 72 nm thick sample after chemical intercalation, revealing a strong gradient in G across the device area. Regions closest to the edges are the most resistive as they have a larger local concentration of Li, compared to regions close to the center that are nearly unlithiated. This is consistent with the picture that the intercalants enter the MoS$_2$ crystal from the edges, and not through the basal plane. Figures 4b and c display vertical and horizontal line scans extracted from the conductance image, showing a contrast of up to 7× between the pristine and lithiated regions. To study the effect of lithium intercalation on morphology of MoS$_2$, we perform AFM on the region highlighted in Fig. 4a. The topographical scan shown in Fig. 4d reveals an excellent correspondence with the conductance map. The pristine region with high thermal conductance is nearly atomically smooth, with a root mean square roughness, $\delta_{RMS}$ ~0.6 nm, while the lithiated region with low thermal conductance shows significantly higher roughness, $\delta_{RMS}$ ~3 nm (Fig. 4e). This region is also thicker by approximately 10 nm, corresponding to a c-axis strain of about 15%. These thermal and morphological features are reproduced in multiple ex situ intercalated MoS$_2$ films (see Supplementary Fig. 6).

**Theoretical modeling.** To understand how Li ions affect the cross-plane thermal conductance of our MoS$_2$ thermal transistors, we perform first principles density functional perturbation theory (DFPT)[28] calculations of phonon dispersion relations. We consider the equilibrium crystal structure 2H-MoS$_2$, with stacking sequence ABAB. Upon intercalation, Li atoms occupy the octahedral sites between the MoS$_2$ layers, forming a thermodynamically unstable 2H-Li$_1$MoS$_2$ phase, which eventually transforms into 1T-Li$_1$MoS$_2$ with stacking sequence AA, and lower in-plane symmetry. We find that the c-axis lattice constant expands by 14% upon intercalation in the 2H-Li$_1$MoS$_2$ phase, but in the stable 1T-Li$_1$MoS$_2$ it is only larger by 0.5%, as compared to 2H-MoS$_2$. The lattice parameters from our calculations (see

Supplementary Table 2) for 2H-MoS$_2$ and 1T-Li$_1$MoS$_2$ agree well with recent measurements[7], which showed a c-axis expansion in bulk Li$_{0.86}$MoS$_2$ of 0.5%. However, previous reports on LiMoS$_2$ powders showed larger expansion (2.5–6%)[26,29]. Our X-ray diffraction (XRD) measurements of electrochemically intercalated Li$_1$MoS$_2$ powders give a c-axis expansion of 2.3% (see Supplementary Fig. 7), while our AFM measurements on chemically lithiated exfoliated single crystals yield a value up to about 15% (Fig. 4e).

We suggest that the relatively large lattice expansion measured in our samples could be due to an incomplete 2H-1T transition, which engenders a mixture of 2H and 1T phases and creates stacking disorder. Previous reports in literature have found evidence of such phase mixtures in lithiated MoS$_2$[7,30]. For example, using in situ Raman spectroscopy in electrochemically intercalated MoS$_2$, Xia et al.[30] observed signatures of the 1T phase remaining after the first recharge (delithiation), even at high electrochemical potentials (2.4 V relative to Li$^+$/Li). They also visualized the coexistence of 2H and 1T phases using high-resolution transmission electron microscopy (HRTEM). The role of stacking disorder has also been suggested by previous ex situ measurements of the thermal conductivity of Li intercalated bulk MoS$_2$. Zhu et al.[7] argued that the strong suppression in the cross-plane thermal conductivity was partly a result of a phase mixture present across the thickness of the crystal. As we will show later using NEMD simulations, this hypothesis of stacking disorder and mixed phases is also consistent with our experimental measurements of c-axis strain and thermal conductance modulation.

Phonon dispersions along the Γ-A direction (Fig. 5a–c) reveal that lithiation gives rise to several flat bands at frequencies above 4 THz in both 2H-Li$_1$MoS$_2$ and 1T-Li$_1$MoS$_2$ (see Supplementary Fig. 8 for a schematic of the Brillouin zone, and dispersions plotted along other high-symmetry directions). The latter phase also displays a large number of MoS$_2$ optical modes, stemming from the larger number of atoms in the unit cell with lower symmetry consisting of four formula units. Phonon modes are color coded to distinguish between those with prevalent MoS$_2$

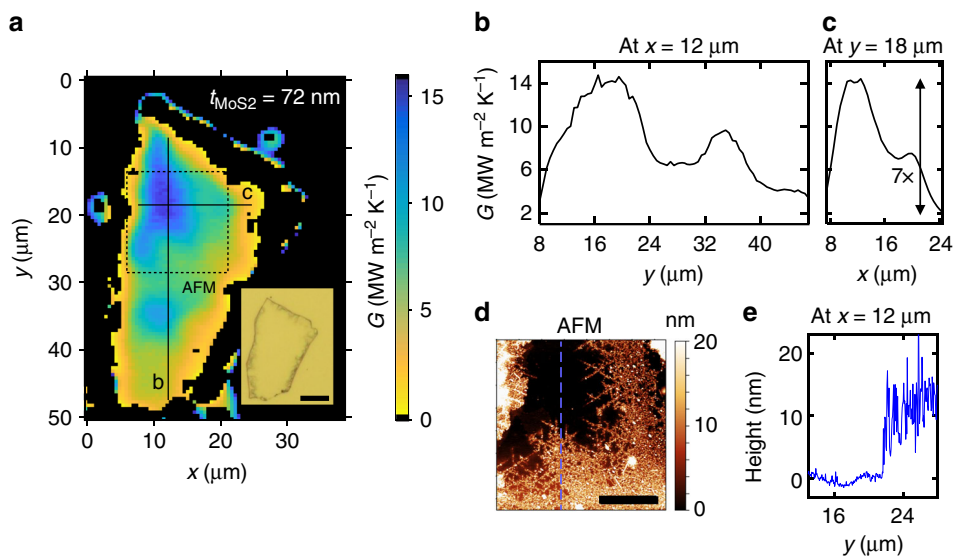

**Fig. 4** Ex situ chemical lithiation experiments. **a** Thermal conductance map of a 72 nm thick single crystal MoS$_2$ film lithiated using n-Butyllithium for 2 h. Inset shows an optical micrograph of the device, after coating with 81 nm thick Al layer (scale bar is 10 μm). **b** Vertical line scan taken at $x = 12$ μm, and **c** horizontal line scan taken at $y = 18$ μm, extracted from the thermal conductance map along the solid lines marked in **a**. **d** AFM image of the region enclosed within the dashed box in **a**, showing a clear correlation between topography and thermal conductance. The smooth pristine region shows the highest thermal conductance, while the rough lithiated region is more thermally resistive. The scale bar is 5 μm. **e** Height profile extracted from the AFM image along the blue dashed line indicated in **d** at $x = 12$ μm

(blue) vs. Li (red) participation. Figures 5d and e show displacement vectors for modes at 6.72 THz in 1T-Li$_1$MoS$_2$ and 4.61 THz in 2H-Li$_1$MoS$_2$, respectively, revealing that Li-related modes are decoupled from the motion of Mo and S atoms. Therefore, we propose that in either crystallographic phase, the Li guest atoms act as rattlers, which reduce phonon lifetimes by increasing the phase space for phonon–phonon scattering[31].

To quantify the impact of Li intercalation on the thermal conductance of thin-film MoS$_2$, we use NEMD. A 10 nm thick film is sandwiched between thermal reservoirs (see Fig. 5f). The cross-plane thermal conductance $G$ is calculated based on the steady-state heat flux, and the temperature gradient developed across the thickness of the film (see Methods, Supplementary Fig. 9, and Supplementary Table 2). Simulations are first performed for 2H-Li$_\chi$MoS$_2$ with the Li composition $\chi$ varying from 0.75 to 1, and the results are plotted in Fig. 5g as a function of the c-axis strain relative to unlithiated 2H-MoS$_2$. The thermal conductance decreases monotonically with increasing $\chi$, consistent with our experimental results. Furthermore, this decrease in $G$ occurs along with a concomitant increase in the c-axis lattice constant. This shows that in addition to increasing the phonon scattering rates, intercalation suppresses $G$ by creating tensile strain along the c-axis, which softens Γ-A phonon modes and reduces group velocities[32]. The maximum reduction in $G$ for 2H-Li$_1$MoS$_2$ is 3.2× relative to the unlithiated 2H-MoS$_2$. While significant, this reduction is less than experimentally measured, suggesting that the large thermal modulation cannot be explained solely by full lithiation of the 2H phase alone. We also consider thermal transport in the fully lithiated 1T phase. As noted earlier, the 1T-Li$_1$MoS$_2$ system has very little c-axis strain (0.5–1%); NEMD calculations of a 10 nm thick film show that the fully intercalated 1T phase shows a comparatively small reduction in $G$ of 1.5×. Furthermore, using ab initio phonon dispersion calculations, we uncover that there is a critical c-axis strain of around 8% beyond which the 1T lithiated phase becomes unstable (see Supplementary Fig. 10). These results suggest that a full phase transition to the 1T-Li$_1$MoS$_2$ phase is unlikely to explain our experimental observations.

As noted above, based on previous reports[7,30] there is a strong likelihood of the existence of mixed 2H and 1T phases and stacking disorder in the lithiated MoS$_2$. In this context, turbostratic disorder has been shown to lead to record-low cross-plane thermal conductivity in layered WSe$_2$ films[33]. To examine the effect of phase mixtures on cross-plane thermal conduction, we construct two systems (each 10 nm thick), which contain stacking disorder along the c-axis. They have the following stacking sequences: (1) {6, 5, 6} layers of {2H, 1T, 2H}, and (2) {4, 3, 4, 2, 4} layers of {2H, 1T, 2H, 1T, 2H}, respectively (see simulation snapshot in Fig. 5f). First, we find that these systems have c-axis strains of 12.5 and 13.5% for Li$_1$MoS$_2$ and Li$_{0.9}$MoS$_2$, respectively, close to the calculated value for 2H-Li$_1$MoS$_2$ (14%). Next, the cross-plane thermal conductance is calculated using NEMD. For Li$_1$MoS$_2$, in both mixed phase systems, we calculate a 7.1× reduction in $G$ compared to the pristine 2H-MoS$_2$. This lowering of $G$ below 2H-Li$_1$MoS$_2$ is likely due to the increased phonon scattering at 2H-1T phase boundaries, qualitatively consistent with the findings of Zhu et al.[7] Furthermore, for the Li$_{0.9}$MoS$_2$ system (i.e., with 10% Li vacancies) we calculate a 9–10× reduction in $G$, i.e., nearly one order of magnitude, comparable to our dynamic TDTR measurements.

## Discussion

A detailed mechanism thus emerges, as indicated in Fig. 5g: as Li intercalates between the MoS$_2$ layers, increased phonon scattering and c-axis tensile strain in the 2H phase lead to a reduction in $G$. With increasing Li concentration, the system transforms partially into the 1T phase, creating 2H-1T phase boundaries along the c-axis. Additional phonon scattering due to stacking disorder in the mixed phase causes a further suppression in $G$, by up to 10× relative to the pristine state. Even in the strongly lithiated state, our calculations and experiments suggest that the thin-film system remains as a phase mixture without undergoing a complete transition to the pure 1T-LiMoS$_2$ state.

We note that an additional source of thermal conductivity suppression is mesoscopic disorder; the high degree of surface

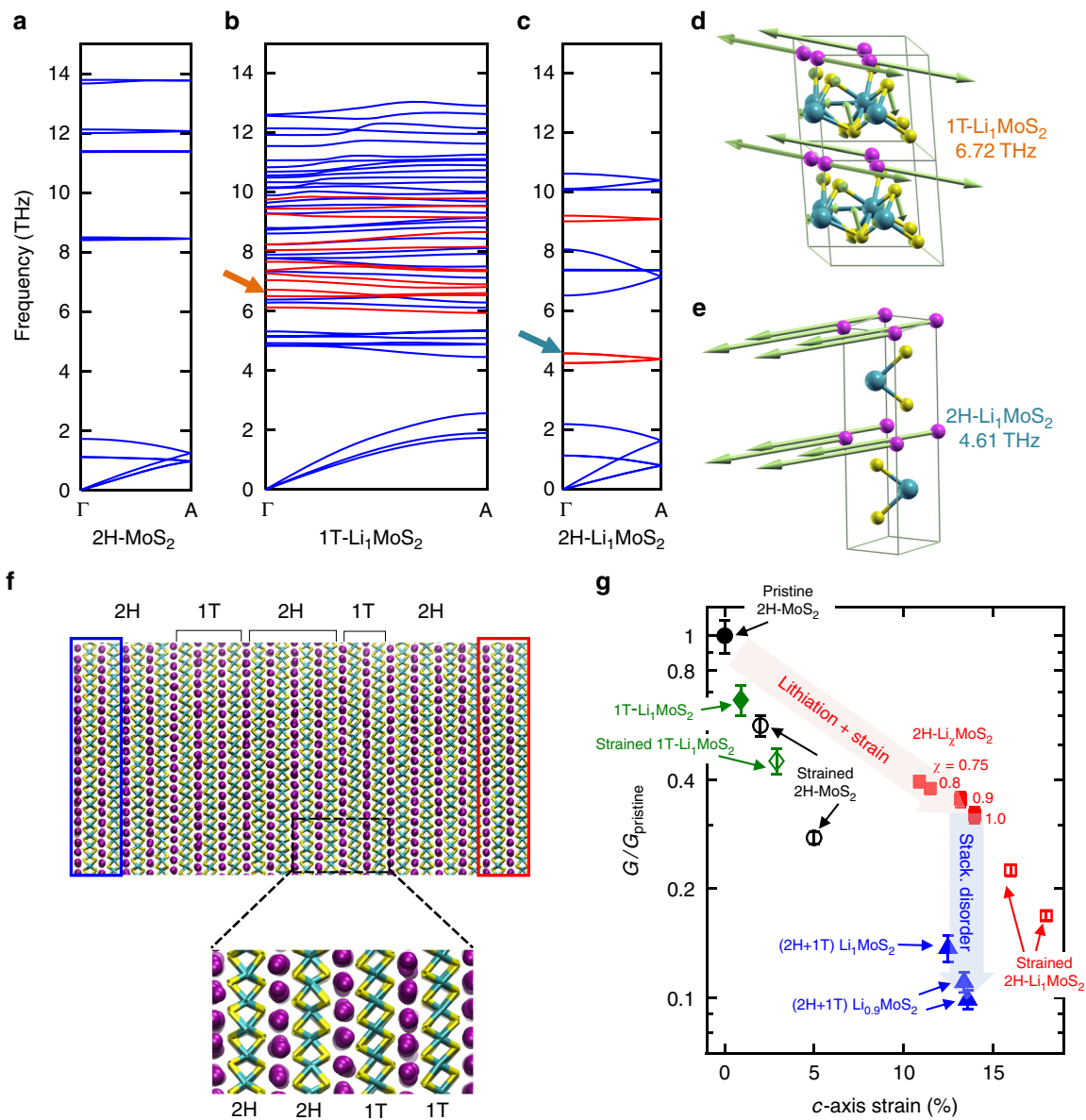

**Fig. 5** Calculated phonon dispersions along the cross-plane Γ-A direction. **a** 2H-MoS$_2$, **b** 1T-Li$_1$MoS$_2$, and **c** 2H-Li$_1$MoS$_2$. Phonon branches are color-coded based on whether they are MoS$_2$-like (blue) or Li-like (red). **d**, **e** Force vectors for the modes at 6.72 THz and 4.61 THz in 1T-Li$_1$MoS$_2$ and 2H-Li$_1$MoS$_2$, respectively, showing strong vibrations of Li atoms (depicted by the magenta spheres). **f** Snapshot of a 10 nm thick NEMD simulation cell showing the mixed phase {4 × 2H, 3 × 1T, 4 × 2H, 2 × 1T, 4 × 2H} Li$_1$MoS$_2$ system, and a zoom-in of a 2H-1T phase boundary. The red and blue boxes are the hot and cold reservoirs, respectively. **g** NEMD calculations of the normalized cross-plane thermal conductance of a 10 nm thick MoS$_2$ film plotted vs. % c-axis strain, relative to the pristine (unlithiated) 2H-MoS$_2$ (shown as the solid black circle). Red solid squares refer to 2H-Li$_\chi$MoS$_2$ with $\chi$ = 0.75, 0.8, 0.9, and 1, blue solid triangles refer to mixed-phase (2H + 1T) Li$_\chi$MoS$_2$ with $\chi$ = 0.9 and 1 (layer sequences for 2H/1T stacking are provided in the main text), and the green solid diamond refers to 1T-Li$_1$MoS$_2$. In each of the above cases, the c-axis strain is intrinsic, i.e., built into the structure because of Li intercalation. The empty symbols—black circles, red squares, and green diamond—refer to externally strained 2H-MoS$_2$, 2H-Li$_1$MoS$_2$, and 1T-Li$_1$MoS$_2$, respectively. Error bars represent statistical uncertainties arising from the fluctuations of the heat current and the temperature profile at stationary non-equilibrium conditions

roughness measured by AFM in the chemically lithiated samples provides some evidence of this (Fig. 4e). Such mesoscopic disorder could result from inhomogeneous strains caused by the non-uniform intercalation of Li ions and the possible formation of microdomains[26]. These would be consistent with the strong inhomogeneities seen in the spatially-resolved TDTR measurements (presented in Fig. 2), as also previous in situ optical microscopy studies of MoS$_2$ flakes undergoing electrochemical intercalation[22].

Finally, we note that since we measure the total thermal conductance of the devices, it is difficult to experimentally decouple the effect of Li on the thermal conductance of the Al/MoS$_2$ and MoS$_2$/SiO$_2$ interfaces. Our ex situ studies indicate that a significant fraction of exfoliated films that are initially adhered to the substrate become detached when lithiated, suggesting that Li weakens the adhesion between MoS$_2$ and SiO$_2$, and thus may reduce the MoS$_2$/SiO$_2$ interfacial conductance. This is consistent with prior calculations of the adhesion energy between graphene and SiO$_2$ with adsorbed interlayers[34], and with measurements of the direct relationship between bond strength and thermal boundary conductance (TBC) in van der Waals junctions[35].

In summary, we have demonstrated an electrochemically driven thermal transistor using Li intercalation in thin MoS$_2$ films. We show large reversible switching of 8–10× in a 10 nm thick

MoS$_2$ device, with an operational time scale of minutes. Using operando scanning thermal conductance microscopy, we reveal inhomogeneities in the distribution of Li within a single MoS$_2$ device, and probe how it evolves over an electrochemical cycle. Correlative AFM imaging provides a direct link between the spatial distribution of Li through its impact on surface topography, and cross-plane thermal transport. Ab initio and NEMD calculations reveal that Li guest atoms act as rattlers within the host MoS$_2$ lattice. The simulations suggest that a combination of multiple factors is likely to lead to such a significant modulation of thermal conductance, including enhanced phonon scattering from Li rattler modes, phonon softening due to lithiation-induced $c$-axis strain, stacking disorder (2H-1T phase mixtures), and mesoscopic disorder.

These results are of crucial significance to the thermal engineering of Li ion batteries, where heat dissipation is a critical issue[36,37]. We have demonstrated that not only does the thermal conductivity of a battery electrode depend on the state-of-charge, it is also spatially inhomogeneous at the microscale. In the context of battery thermal modeling, we suggest that it may be important to consider dynamic spatio-temporal variations in thermal conductivity within single electrode particles. In other words, low thermal conductivity regions could be linked to the formation of hot spots, and can accelerate eventual thermal runaway of the battery.

In addition, as an operando technique, we contend that thermal conductance microscopy can provide useful information about changes in the local microstructure of nanomaterials that are undergoing reactions. Our approach leverages the strong sensitivity of phonon scattering rates to local defect density and strain, and represents a fundamentally different mechanism for microscopy of dynamically evolving materials. Furthermore, as an optical technique, it is complementary to existing electron or X-ray based techniques, which can often require specialized sample preparation.

In the context of active thermal management, our thermal transistor results represent an important advance with their nearly one order of magnitude switching ratio. For example, in power electronics where thermal cycling can lead to catastrophic failure due to thermal expansion, the inclusion of a thermal transistor between the device and heat sink can greatly suppress temperature oscillations and improve reliability. Our calculations suggest that to first order, a thermal transistor with 10× on/off ratio can reduce the temperature swing in a device by 10×, and increase its lifetime by almost 3000×. This is related to a modulation of the device's cooling time constant, $\tau \sim C_{thermal}/G$, by switching $G$ to a lower value in the off-state. The temperature swing, $\Delta T$, is proportional to $[1 - \exp(-t_{off}/\tau)] \sim t_{off}/\tau$, where $t_{off}$ is the time during which the device is off. Reducing the temperature swing can have a significant effect on device reliability as the relationship between $\Delta T$ and the number of cycles to failure $N_f$ is strongly non-linear[38], $N_f \sim \Delta T^{-3,5}$.

We note that for electrochemical thermal transistors to reach deployment in technological applications, several challenges remain to be overcome. First, it will be important to understand the impact of device thickness on the thermal on/off ratio, and therefore decouple the effect of intrinsic (volumetric) vs. interface thermal switching. From a practical standpoint, in our experiments we found it challenging to perform electrochemical measurements on films thicker than tens of nanometers, likely due to lithiation-induced volume changes and consequent electrode detachment. Furthermore, to ensure robust switching under different operating conditions, it will be necessary to implement a fully solid-state version of this device. This could potentially be achieved using solid polymer electrolytes which have seen widespread adoption in Li ion batteries[39]. Lastly, efforts must be made to increase thermal switching speeds in order to respond better to the fast heat flux transients typically found in electronic devices. Since the time scales for Li ion diffusion scale quadratically with length scale, it could be promising to engineer (arrays of) thermal switches with reduced lateral features.

Besides active thermal management and heat routing, our results also have direct implications for energy harvesting. For example, a pyroelectric device placed in contact with a thermal transistor can be used to harvest electrical energy from a passive, time-invariant heat source. This would be achieved by externally modulating the transistor's thermal conductance to generate temporal variations in the pyroelectric material's temperature. Additionally, intercalated layered materials also hold immense promise for thermoelectric applications. Previous studies have shown that Li intercalation enhances the in-plane electrical conductivity of MoS$_2$ and other layered materials by as much as two orders of magnitude[22,40]. While there have not been, to the best of our knowledge, measurements of cross-plane electrical transport in lithiated MoS$_2$, we anticipate a similar enhancement due to increased carrier density. Intercalation thus potentially offers a unique mechanism to achieve order of magnitude improvements in the thermoelectric figure of merit $zT$ for atomically-thin energy conversion devices of the future.

## Methods

**Fabrication of operando electrochemical platform.** Crystalline, layered MoS$_2$ films were mechanically exfoliated onto SiO$_2$ (90 nm) on p-type Si substrates. Optical microscopy was used to select an appropriate device, and AFM was used to measure its thickness and confirm uniformity. An 80 nm thick Al layer was selectively patterned on top of the MoS$_2$ using electron-beam (e-beam) lithography and e-beam evaporation, while leaving the edges exposed to allow Li ions to intercalate. A 50 nm thick Cu electrode was e-beam evaporated through a shadow mask, patterned approximately 1 cm away from the MoS$_2$ device. The sample was transferred into an Ar-filled glove box, where a small Li metal pellet was placed onto the Cu electrode. A glass cover slip (around 0.2 mm thick) was placed on top of the MoS$_2$ device and Li metal, and sealed on three sides using epoxy. The pocket was filled with electrolyte (1.0 M LiPF$_6$ in EC/DEC, 1:1 w/w) and the fourth side was sealed with epoxy to prevent oxidation. Care was taken to minimize bubble formation in the electrolyte, which could disrupt the optical measurements. Wire bonds were used to make electrical contact to the terminals of the potentiostat (Gamry Instruments Reference 600) for electrochemical measurements. This was done while taking precautions to prevent electrostatic discharge, which was one of the major factors affecting device yield.

**Time-domain thermoreflectance.** Cross-plane thermal transport in MoS$_2$ thin films was measured using TDTR, an optical pump-probe technique. Details of this technique and our setup are provided elsewhere[11,12,41]. In these experiments, we used a pump modulation frequency of 4 MHz, and pump and probe 1/e$^2$ spot diameters of 4.0 μm and 2.7 μm, respectively. The transient temperature variation created by the absorption of modulated pump pulses within the Al transducer was measured by monitoring the reflected probe intensity, which was demodulated using a radio-frequency lock-in amplifier (Stanford Research Systems 844). The signal consists of the ratio of the in-phase ($V_{in}$) to out-of-phase ($V_{out}$) voltage (= −$V_{in}/V_{out}$) measured by the lock-in as a function of pump-probe delay time, which was varied from 0 to 3.6 ns using a mechanical delay stage. These data were fit to a three-dimensional multilayer heat diffusion model that accounts for bidirectional heat flow, radial heat spreading due to the finite spot size, anisotropic thermal conductivity of MoS$_2$, and TBC between adjacent materials[42–44]. For operando measurements performed through the transparent electrolyte, the total optical power was limited to ~3 mW to prevent laser-induced degradation of the Al surface. The estimated steady-state temperature rise was <2 K. An integrated dark-field microscope enabled location and imaging of MoS$_2$ thermal transistor devices.

Real-time operando measurements were performed by fixing the probe delay time at +100 ps and collecting the TDTR ratio signal. The temporal resolution is limited by the time-constant of the lock-in amplifier (100 ms), which is significantly faster than intercalation time scales in our device (minutes). For the single-spot time-dependent measurements shown in Fig. 3, $V_{in}$ and $V_{out}$ were recorded every 1 s while the device was subjected to repeated cycles of galvanostatic charge and discharge (see Supplementary Fig. 11 for raw TDTR data). Similar measurements on a control device, i.e., Al/SiO$_2$/Si, showed no changes in TDTR signal with electrochemical potential, ruling out intercalation-induced changes in the properties of the Al transducer and electrolyte/Al interface. Measured ratio (= −$V_{in}/V_{out}$) data at the +100 ps delay time were converted to thermal conductance $G$ by comparing with a correlation curve calculated from the multilayer thermal model (see Supplementary Fig. 3, Supplementary Table 1). Thermophysical properties of the liquid electrolyte were measured using a through-substrate TDTR approach (see Supplementary Fig. 2).

Spatial thermal conductance imaging using TDTR was performed by mounting the sample on a motorized stage (Melles Griot Nanomotion II), which has a resolution of 10 nm and a bidirectional repeatability of 100 nm. After holding the $MoS_2$ device at a constant $V_{WE}$ (relative to $Li^+/Li$) for about 10 min, the sample was raster scanned in the plane normal to the laser beam while recording $V_{in}$ and $V_{out}$ at a fixed delay time of +100 ps, with a step size of 500 nm and dwell time of 300 ms per pixel (see Fig. 2b). A typical 40 × 40 μm scan takes around 30 min. Pixels near the edges of the device give unphysical thermal conductance values due to optical edge effects (and for ex situ samples, also due to oxidation), and are not plotted on the colormap. To define valid data points, we used the following criterion: $0.9\ V_{in,0} < V_{in} < 1.1\ V_{in,0}$, where $V_{in,0}$ is the mean $V_{in}$ value inside the device. This is based on the fact that at short delay times, $V_{in}$ is not significantly sensitive to $G$, so that large variations in $V_{in}$ indicate abrupt changes in device reflectivity. Measured TDTR ratio values at each pixel were converted to $G$ using the procedure mentioned above.

In the operando measurements, we observed that optical propagation through the liquid electrolyte causes an asymmetry in the $V_{out}$ signal as a function of distance $z$ on either side of the focal plane. This causes the TDTR ratio signal to be asymmetric as well, which could introduce an error in the measurements if the sample is not at the correct $z$ plane. We posit that this occurs due to a thermo-optic modulation of the refractive index of a thin layer of liquid due to heat conducted away from the Al transducer. This effect (which occurs at relatively long time scales due to the low thermal conductivity of the electrolyte) likely leads to a modification of the effective thermoreflectance coefficient of the metal by a scaling factor, affecting the $V_{out}$ signal component. To correct for this, we scale the TDTR ratio data by a small factor (1.05), calibrating such that the thermal conductance of the device in the unlithiated state under liquid be equal to that of the pristine device measured prior to liquid encapsulation (Supplementary Fig. 1). This correction factor does not significantly change the measured thermal on/off ratio between the unlithiated and lithiated states.

**Ab initio and molecular dynamics calculations**. The first principles DFT calculations were performed in local density approximation (LDA) of the exchange and correlation functional[45]. Core electrons were approximated using norm-conserving pseudopotentials[46], and the Kohn–Sham wavefunctions were expanded on a plane wave basis set with a cutoff of 100 Rydberg (1360 eV). Integration of the electronic properties over the first Brillouin zone was performed using the following Monkhorst–Pack meshes of $k$-points:[47] 10 × 10 × 4 for 2H-$MoS_2$ and 2H-$Li_1MoS_2$, and 4 × 4 × 4 for 1T-$Li_1MoS_2$. Structural and cell relaxations were performed by the Broyden-Fletcher-Goldfarb-Shanno (BFGS) quasi-Newton algorithm with a strict convergence criterion of 1E-8 Rydberg/Bohr for maximum residual force component. Phonon dispersion relations were computed by density-functional perturbation theory (DFPT)[28], with 10 × 10 × 4, 4 × 4 × 4, and 4 × 4 × 2 $q$-mesh for pure 2H-$MoS_2$, 1T-$Li_1MoS_2$, and 2H-$Li_1MoS_2$, respectively. All the calculations were performed using the Quantum-Espresso package[48].

From the physical properties computed by DFT (lattice parameters, phonon dispersion relations, speed of sound) we fitted the parameters of a two-body empirical potential for NEMD calculations. The functional form of the potential consists of a combination of Lennard–Jones (LJ) interactions and electrostatics: $V(r_{ij}) = 4\varepsilon_{ij}\left[\left(\frac{\sigma_{ij}}{r_{ij}}\right)^{12} - \left(\frac{\sigma_{ij}}{r_{ij}}\right)^{6}\right] + \frac{Q_iQ_j}{r_{ij}}$. Partial charges ($Q$) and LJ parameters ($\sigma$ and $\varepsilon$) are provided in Supplementary Table 2. This potential was used to perform NEMD simulations of thermal transport employing the reverse-NEMD approach[49], in a nano-device of the same thickness as the experimental one. Supercells consisting of a total of 34 $MoS_2$ layers were used, in which two equally spaced two-layer slabs operate as hot and cold thermal baths. The system has a lateral size of 2.76 nm by 2.55 nm, and periodic boundary conditions are applied in all directions. The systems were first equilibrated in the weak-coupling NPT ensemble for 200 ps, and eventually in the NVT ensemble for further 200 ps. In reverse-NEMD, a stationary heat current ($J$) is set by exchanging the momentum of a particle with high kinetic energy in the cold thermal bath, with the momentum of a particle with low kinetic energy in the hot thermal bath. The stationary value of the heat current is determined by the momentum exchange rate: our simulations were carried out for 4 ns with an exchange rate of 1000 time steps (each MD time step is 1 fs). To compute thermal conductance, one could directly use Kapitza's definition $G = J/\Delta T$ under stationary conditions. However, to avoid spurious contact effects, we instead computed the thermal conductivity ($\kappa$) of the finite system using Fourier's law $J = \kappa \nabla T$, and used $G = \kappa/L$, where $L$ is the distance between the thermal baths. Temperature profiles and heat flux as a function of time and are shown in Supplementary Fig. 9.

**Data availability:**
The data that support the findings of this study are available from the corresponding authors upon reasonable request.

## References

1. Sood, A., Pop, E., Asheghi, M. & Goodson, K. E. The heat conduction renaissance. In *17th IEEE Intersociety Conference on Thermal and Thermomechanical Phenomena in Electronic Systems (ITherm)* 1396–1402 (IEEE, 2018).
2. Yan, Y. & Malen, J. A. Periodic heating amplifies the efficiency of thermoelectric energy conversion. *Energy Environ. Sci.* **6**, 1267–1273 (2013).
3. Wang, L. & Li, B. Thermal logic gates: computation with phonons. *Phys. Rev. Lett.* **99**, 177208 (2007).
4. Li, N. et al. Colloquium: phononics: manipulating heat flow with electronic analogs and beyond. *Rev. Mod. Phys.* **84**, 1045–1066 (2012).
5. Walker, C. T. & Pohl, R. O. Phonon scattering by point defects. *Phys. Rev.* **131**, 1433–1442 (1963).
6. Callaway, J. & von Baeyer, H. C. Effect of point imperfections on lattice thermal conductivity. *Phys. Rev.* **120**, 1149–1154 (1960).
7. Zhu, G. et al. Tuning thermal conductivity in molybdenum disulfide by electrochemical intercalation. *Nat. Commun.* **7**, 13211 (2016).
8. Klemens, P. G. Phonon scattering by oxygen vacancies in ceramics. *Phys. B Condens. Matter* **263–264**, 102–104 (1999).
9. Sproull, R. L., Moss, M. & Weinstock, H. Effect of dislocations on the thermal conductivity of lithium fluoride. *J. Appl. Phys.* **30**, 334–337 (1959).
10. Wang, Z., Alaniz, J. E., Jang, W., Garay, J. E. & Dames, C. Thermal conductivity of nanocrystalline silicon: Importance of grain size and frequency-dependent mean free paths. *Nano. Lett.* **11**, 2206–2213 (2011).
11. Sood, A. et al. Anisotropic and inhomogeneous thermal conduction in suspended thin-film polycrystalline diamond. *J. Appl. Phys.* **119**, 175103 (2016).
12. Sood, A. et al. Direct visualization of thermal conductivity suppression due to enhanced phonon scattering near individual grain boundaries. *Nano. Lett.* **18**, 3466–3472 (2018).
13. Costescu, R. M., Cahill, D. G., Fabreguette, F. H., Sechrist, Z. A. & George, S. M. Ultra-low thermal conductivity in $W/Al_2O_3$ nanolaminates. *Science* **303**, 989–990 (2004).
14. Lee, J. et al. Phonon and electron transport through $Ge_2Sb_2Te_5$ films and interfaces bounded by metals. *Appl. Phys. Lett.* **102**, 191911 (2013).
15. Yang, J. et al. Enhanced and switchable nanoscale thermal conduction due to van der Waals interfaces. *Nat. Nanotechnol.* **7**, 91–95 (2012).
16. Wehmeyer, G., Yabuki, T., Monachon, C., Wu, J. & Dames, C. Thermal diodes, regulators, and switches: physical mechanisms and potential applications. *Appl. Phys. Rev.* **4**, 041304 (2017).
17. Oh, D. W., Ko, C., Ramanathan, S. & Cahill, D. G. Thermal conductivity and dynamic heat capacity across the metal-insulator transition in thin film $VO_2$. *Appl. Phys. Lett.* **96**, 151906 (2010).
18. Zheng, R., Gao, J., Wang, J. & Chen, G. Reversible temperature regulation of electrical and thermal conductivity using liquid-solid phase transitions. *Nat. Commun.* **2**, 289 (2011).
19. Ihlefeld, J. F. et al. Room-temperature voltage tunable phonon thermal conductivity via reconfigurable interfaces in ferroelectric thin films. *Nano. Lett.* **15**, 1791–1795 (2015).
20. Cho, J. et al. Electrochemically tunable thermal conductivity of lithium cobalt oxide. *Nat. Commun.* **5**, 4035 (2014).
21. Kang, J. S., Ke, M. & Hu, Y. Ionic intercalation in two-dimensional van der Waals materials: In-situ characterization and electrochemical control of the anisotropic thermal conductivity of black phosphorus. *Nano. Lett.* **17**, 1431–1438 (2017).
22. Xiong, F. et al. Li intercalation in $MoS_2$: in situ observation of its dynamics and tuning optical and electrical properties. *Nano. Lett.* **15**, 6777–6784 (2015).
23. Lim, J. et al. Origin and hysteresis of lithium compositional spatiodynamics within battery primary particles. *Science* **353**, 566–571 (2016).
24. Wan, J. et al. In situ investigations of Li-$MoS_2$ with planar batteries. *Adv. Energy Mater.* **5**, 1401742 (2015).
25. Dreyer, W. et al. The thermodynamic origin of hysteresis in insertion batteries. *Nat. Mater.* **9**, 448–453 (2010).
26. Py, M. A. & Haering, R. R. Structural destabilization induced by lithium intercalation in $MoS_2$ and related compounds. *Can. J. Phys.* **61**, 76–84 (1983).
27. Wang, H. et al. Electrochemical tuning of vertically aligned $MoS_2$ nanofilms and its application in improving hydrogen evolution reaction. *Proc. Natl Acad. Sci. USA* **110**, 19701–19706 (2013).
28. Baroni, S., De Gironcoli, S., Dal Corso, A. & Giannozzi, P. Phonons and related crystal properties from density-functional perturbation theory. *Rev. Mod. Phys.* **73**, 515–562 (2001).

29. Mulhern, P. J. Lithium intercalation in crystalline Li$_x$MoS$_2$. *Can. J. Phys.* **67**, 1049–1052 (1989).
30. Xia, J. et al. Phase evolution of lithium intercalation dynamics in 2H-MoS$_2$. *Nanoscale* **9**, 7533–7540 (2017).
31. Tadano, T., Gohda, Y. & Tsuneyuki, S. Impact of rattlers on thermal conductivity of a thermoelectric clathrate: a first-principles study. *Phys. Rev. Lett.* **114**, 095501 (2015).
32. Ding, Z., Jiang, J.-W., Pei, Q.-X. & Zhang, Y.-W. In-plane and cross-plane thermal conductivities of molybdenum disulfide. *Nanotechnology* **26**, 065703 (2015).
33. Chiritescu, C. et al. Ultralow thermal conductivity in disordered, layered WSe$_2$ crystals. *Science* **315**, 351–353 (2007).
34. Gao, W., Xiao, P., Henkelman, G., Liechti, K. M. & Huang, R. Interfacial adhesion between graphene and silicon dioxide by density functional theory with van der Waals corrections. *J. Phys. D. Appl. Phys.* **47**, 255301 (2014).
35. Losego, M. D., Grady, M. E., Sottos, N. R., Cahill, D. G. & Braun, P. V. Effects of chemical bonding on heat transport across interfaces. *Nat. Mater.* **11**, 502–506 (2012).
36. Finegan, D. P. et al. In-operando high-speed tomography of lithium-ion batteries during thermal runaway. *Nat. Commun.* **6**, 6924 (2015).
37. Bandhauer, T. M., Garimella, S. & Fuller, T. F. A critical review of thermal issues in lithium-ion batteries. *J. Electrochem. Soc.* **158**, R1–R25 (2011).
38. Bayerer, R., Herrmann, T., Licht, T., Lutz, J. & Feller, M. Model for power cycling lifetime of IGBT modules–various factors influencing lifetime. In *5th International Conference on Integrated Power Systems (CIPS)* 1–6 (VDE, 2008).
39. Yue, L. et al. All solid-state polymer electrolytes for high-performance lithium ion batteries. *Energy Storage Mater.* **5**, 139–164 (2016).
40. Sharbati, M. T. et al. Low-power, electrochemically tunable graphene synapses for neuromorphic computing. *Adv. Mater.* **30**, 1802353 (2018).
41. Sood, A. et al. Thermal conduction in lattice-matched superlattices of InGaAs/InAlAs. *Appl. Phys. Lett.* **105**, 051909 (2014).
42. Cahill, D. G. Analysis of heat flow in layered structures for time-domain thermoreflectance. *Rev. Sci. Instrum.* **75**, 5119–5122 (2004).
43. Schmidt, A., Chiesa, M., Chen, X. & Chen, G. An optical pump-probe technique for measuring the thermal conductivity of liquids. *Rev. Sci. Instrum.* **79**, 064902 (2008).
44. Schmidt, A. J., Chen, X. & Chen, G. Pulse accumulation, radial heat conduction, and anisotropic thermal conductivity in pump-probe transient thermoreflectance. *Rev. Sci. Instrum.* **79**, 114902 (2008).
45. Perdew, J. P. & Zunger, A. Self-interaction correction to density-functional approximations for many-electron systems. *Phys. Rev. B* **23**, 5048–5079 (1981).
46. Hartwigsen, C., Goedecker, S. & Hutter, J. Relativistic separable dual-space Gaussian pseudopotentials from H to Rn. *Phys. Rev. B* **58**, 3641–3662 (1998).
47. Monkhorst, H. J. & Pack, J. D. Special points for Brillouin-zone integrations. *Phys. Rev. B* **13**, 5188–5192 (1976).
48. Giannozzi, P. et al. Quantum Espresso: a modular and open-source software project for quantum simulations of materials. *J. Phys. Condens. Matter* **21**, 395502 (2009).
49. Müller-Plathe, F. A simple nonequilibrium molecular dynamics method for calculating the thermal conductivity. *J. Chem. Phys.* **106**, 6082–6085 (1997).

### Acknowledgements
We thank Ramez Cheaito, Chi Zhang, Woosung Park, and Eilam Yalon of Stanford University for helpful discussions. We acknowledge the Stanford Nanofabrication Facility (SNF) and Stanford Nano Shared Facilities (SNSF) for enabling device fabrication and measurements, funded under National Science Foundation (NSF) award ECCS-1542152. This work was supported in part by the NSF Engineering Research Center for Power Optimization of Electro Thermal Systems (POETS) with cooperative agreement EEC-1449548, by NSF EFRI 2-DARE grant 1542883, by AFOSR grant FA9550-14-1-0251, and by the Stanford SystemX Alliance. F.X. and Y.C. were partially supported by the Department of Energy, Office of Basic Energy Sciences, Division of Materials Science and Engineering, under contract DE-AC02-76SF00515.

### Author contributions
F.X., Y.C., and E.P. proposed the idea of measuring the impact of Li on MoS$_2$ thermal transport. A.S. and K.E.G. conceived the in situ spatio-temporal thermal transport experiments. F.X. fabricated the devices with assistance from H.W., J.Z., C.M., and J.S. A.S. developed the thermal conductance microscopy technique and performed TDTR experiments, electrochemical measurements, correlative AFM characterization, and analyzed all experimental data, with advice and input from K.E.G. D.D. designed modeling, S.C. performed DFT calculations, D.D. and D.S. performed MD simulations. A.S. synthesized the experimental and theoretical results and wrote the manuscript draft, with inputs from D.D. and E.P. Y.C., E.P., and K.E.G. supervised the project. All authors discussed the results, commented on the manuscript, and approved the final version.

### Additional information
**Supplementary Information** accompanies this paper at https://doi.org/10.1038/s41467-018-06760-7.

**Competing interests:** The authors declare no competing interests.

**Reprints and permission** information is available online at http://npg.nature.com/reprintsandpermissions/

**Publisher's note:** Springer Nature remains neutral with regard to jurisdictional claims in published maps and institutional affiliations.